\documentclass[eqsecnum,floats,aps,footinbib,prd,twocolumn,floatfix,showpacs]{revtex4-1} 
\usepackage{latexsym,pifont,textcomp,setspace,xspace}
\usepackage{amsmath,amstext,amssymb,color,graphicx}
\usepackage{bm}
\usepackage[utf8]{inputenc}
\usepackage{soul} 

\usepackage{amsmath}
\usepackage{amssymb}
\usepackage{indentfirst}
\everymath{\displaystyle}
\usepackage{graphicx}
\usepackage{setspace}

\usepackage{color}

\newcommand{\PC}[1]{\ensuremath{\left(#1\right)}}

\usepackage[english]{babel}

\definecolor{purple}{rgb}{1,0,1}

\allowdisplaybreaks
\begin{document}
\title{Tolman temperature gradients in a gravitational field}
\author{Jessica Santiago and Matt Visser\vspace{3pt}}
\affiliation{School of Mathematics and Statistics,
Victoria University of Wellington; \\
PO Box 600, Wellington 6140, New Zealand.}
\begin{abstract}
Tolman's relation for the temperature gradient in an equilibrium self-gravitating general relativistic fluid is broadly accepted within the general relativity community. 
However, the concept of temperature gradients in thermal equilibrium continues to cause confusion in other branches of physics, since it contradicts \emph{naive} versions of the laws of classical thermodynamics. 
In this paper we  discuss the crucial role of the universality of free fall, and how
thermodynamics emphasises the great distinction between gravity and other forces.
To do so we will present an argument given by Maxwell and apply it to an electro-thermal system, concluding with an \emph{reductio ad absurdum}.
Among other issues we shall show that Tolman temperature gradients could also (in principle) have been derived \emph{circa} 1905 --- a decade before the development of full general relativity. 
		
\medskip\noindent
\emph{Date:} 12 March  2018, LaTeX-ed \today.

%

\end{abstract}
\pacs{04.20.-q;  04.40.-b; 05.20.-y; 05.70.-a}
\maketitle
\def\eref#1{(\ref{#1})}

\section{Introduction}
The \emph{naive} version of the zeroth law of thermodynamics states that if a system $\Sigma$ is in thermal equilibrium, then the temperature will be constant throughout all points $p \in \Sigma$. In another words, the equilibrium temperature field is spatially and temporally uniform as long as no external changes are applied. Furthermore,  the \emph{naive} version of the second law of thermodynamics, states (in Clausius form) that if two systems with different temperatures are put into thermal contact with each other, heat will flow from the hotter to the colder body until thermal equilibrium is reached. Both of these thermodynamic laws need subtle rephrasing in the presence of gravity --- instead of the locally measured temperature $T(x)$ being the determining factor, it is the redshifted temperature~\cite{tolman:1930}
\begin{equation}
T_0  = T(z) \; \left\{ 1 + {g z\over c^2} \right\}
\label{E:T0}
\end{equation}
that is internally constant in equilibrium (zeroth law), and it is differences in $T_0$, not differences in the locally measured $T(x)$, that drive the direction of heat flow in the Clausius form of the second law~\cite{footnote:0}.

Consider a finite-size box filled with some fluid, and assume it to be placed in a gravitational field. We might think of this system as being composed of several subsystems, for example, layers at distinct heights where all the individual layers are in thermodynamical contact with their surroundings through permeable diathermic walls. When the state of thermal equilibrium is reached, we expect, according to the \emph{naive} laws of thermodynamics, to have a spatially and temporally uniform temperature through the whole box.

\vspace{100pt}
On the other hand, in 1930, Richard Tolman wrote an enlightening paper~\cite{tolman:1930} arguing that heat, as much as any other source of energy, is subjected to gravity. Tolman's conclusion, after calculations with both massive and radiation gases, is that the equilibrium temperature of a spherically symmetric static distribution of a perfect fluid is not uniform. It depends, given the symmetry of the spacetime, and using $(-+++)$ metric signature, only on the radial position: 
\begin{equation}
T(r)= {T_0 \over\sqrt{-g_{00}(r)}}; \qquad T_0 = T(r) \sqrt{-g_{00}(r)}.
\label{E:T1}
\end{equation}
Here $T_0$ is simply the integration constant that physically corresponds to the temperature seen by an observer at $r=\infty$, and $g_{\mu\nu}$ are the metric components. So, accordingly to Tolman's results, a column of gas situated in a gravitational field will have a temperature gradient depending on the height even after thermal equilibrium is reached. 
In the weak-field flat-Earth approximation
\begin{equation}
g_{00} \approx -1 -2\Phi \approx -1 - {2gz\over c^2},
\end{equation}
in agreement with equation \eref{E:T0}.
Near the surface of the Earth, $\nabla T(z)/T(z) \approx 10^{-16} \mathrm{m}^{-1}$, which is negligible in most experimental settings.  

\enlargethispage{20pt}
Conversely, spherical symmetry can easily be dispensed with and a modified version of Tolman's relation \eref{E:T1} also holds for arbitrary static spacetimes, as long as the fluid is also static, and
as long as one adopts the  time coordinate for which the metric is manifestly time independent and block diagonal.
We then have, in a completely general spatial-position-dependent manner~\cite{ehrenfest:1930}
\begin{equation}
T(x)= {T_0 \over\sqrt{-g_{00}(x)}}; \qquad T_0 = T(x) \sqrt{-g_{00}(x)}.
\label{E:T2}
\end{equation}

\clearpage
Nothing said up to this point is really news for members of the general relativity community, although scientists from other areas of physics still largely ignore the existence of Tolman's results. Despite being already 88 years old, Tolman's temperature gradient result, combined with classical thermodynamics, is still able to provide useful information emphasizing the great differences between gravity and other forces of nature.

For instance Tolman's result necessarily implies a subtle modification of Fick's law for heat fluxes; Fick's law must really be rephrased in terms of the redshifted temperature $T_0$, not the locally measured temperature $T(x)$:
\begin{equation}
\hbox{(heat flux)} \propto \nabla T_0 = \nabla \left( T(x) \sqrt{-g_{00}(x)} \right). 
\end{equation}
Only with this modified Fick's law will there be no heat fluxes in thermal equilibrium.

This paper is organized as follows: In section \ref{S:tolman} we will briefly recapitulate some of Tolman's results, pointing out some crucial relations for a radiation gas. 
Section \ref{S:radiation} extends Tolman's results to a special relativistic radiation gas subject to Newtonian gravity --- strictly speaking \emph{general} relativity is not needed. 
We will also look at the Planck spectrum in section~\ref{S:planck},
and in section \ref{S:temperature} develop a distinct interpretation of what is temperature by asking what an external observer at a fixed position will see. Next in section \ref{S:maxwell} we will discuss an argument given by Maxwell regarding  thermal equilibrium for a column of gas in a gravitational field. In Section \ref{S:electric} we study the thermodynamic consequences of a mooted temperature gradient induced by electric fields; quickly leading to a \emph{reductio ad absurdum}. 
Finally Section \ref{S:conclusions} discusses our conclusions in this present work.
(In two appendices we first discuss how history might have been different --- Tolman-like results could in principle have been obtained as early as 1875, and second we present some brief comments on Loschmidt's argument.)

\section{Tolman's argument}\label{S:tolman}
Let us now take a brief moment to review the main points of Tolman's result \cite{tolman:1930}. Taking the energy-momentum tensor of a perfect fluid,
\begin{equation}
T^{a b} = (\rho + p)\;u^a u^b + p g^{ab},
\end{equation}
with $p$ being the isotropic pressure of the fluid, and $\rho$ its density, plus a general static spherically symmetric spacetime of the form~\cite{footnote:1}
\begin{equation}
ds^2 = -e^{\nu(r)}dt^2 + e^{\mu(r)}\PC{dr^2 + r^2 d\Omega^2},
\end{equation}
we can impose both Einstein's equation as well as the fluid continuity relation. Doing so, we obtain the following result, 
equivalent to the general relativistic Euler equation:
\begin{equation}
\label{E:euler}
\frac{\partial p}{\partial r} = -\frac{\rho +p}{2}\;  \frac{\partial \nu}{\partial r},
\end{equation}

When dealing with black body radiation, it is easy to see that applying the Stefan--Boltzmann law, $\rho = a\, T^4$, together with the equation of state $p = (\rho/3)$ into equation \eqref{E:euler}, we arrive at:
\begin{equation}
\frac{d \ln T}{dr} = -\frac{1}{2}\;\frac{d\nu}{dr},
\end{equation}
which leads to the temperature dependency on the metric 
\begin{equation}
T(r) = T_0\; e^{-\nu(r) /2}.
\end{equation}
Here $T_0$ is an integration constant that physically corresponds to the temperature seen by an observer at $r=\infty$, (where we take $\nu(\infty)=0$). 

With hindsight, the key step in the argument above is the covariant conservation of stress-energy, leading to the general-relativistic Euler equation. 
The Einstein equations are actually a side issue --- after all in any realistic system there may be many other forms of stress-energy contributing to generating the gravitational field. As long as the radiation fluid has a stress-energy tensor that is independently conserved, then Tolman's result will follow. (This particular issue is partly addressed in the slightly later Tolman--Ehrenfest article~\cite{ehrenfest:1930}.)

Now, taking a closer look to the general case of a fluid consisting of massive particles, although the relativistic Euler equation \eqref{E:euler} remains valid, it is now necessary to have some extra information to connect the fluid variables $(\rho, p)$ to the temperature. Tolman did this by implementing a  generalized (covariant) version of the second law of thermodynamics~\cite{tolman:1928}, which he then applied in his 1930 article~\cite{tolman:1930}. (There are also other ways of dealing with a massive fluid, \emph{vide} Maxwell's argument discussed below, or the Tolman--Ehrenfest article~\cite{ehrenfest:1930}.
See also~\cite{tolman:1933a,tolman:1933b}.)

For a general static metric of the form
\begin{equation}
ds^2 = -g_{00}\; dt^2 + g_{ij}\;dx^i dx^j,
\end{equation}
Tolman's temperature gradient reads~\cite{ehrenfest:1930}:
\begin{equation}
\label{E:tolman temp}
T(x) = {T_0 \over \sqrt{-g_{00}(x)}}; \qquad  T_0 = T(x) \, \sqrt{-g_{00}(x)}.
\end{equation}

\section{Classical radiation gas}\label{S:radiation}

In this section we will argue that the Tolman temperature gradient could have been derived some 25 years earlier than it actually was, in 1905 instead of 1930.  

Indeed the Tolman temperature gradient could have been derived once special relativity was developed, the key choke points being $E=mc^2$ and the relativistic Euler equation.
Already in the mid 1800's the notion of radiation gas was common, this being a gas of electro-magnetic wave-packets, (the notion of photon then not yet being developed),  with energy density and pressure determined by Stefan's radiation law: $dP = \sigma T^4 dA$. When phrased in terms of the radiation constant $a$ we have
\begin{equation}
u = a T^4; \qquad p = {1\over3} a T^4.
\end{equation}

The second step, really only available once one has the notion of $E=mc^2$ firmly under control, is to realise the radiation gas has a equivalent mass density $\rho = u/c^2$. 

The third step involves invoking the special relativistic Euler equation (with $g$ being the local Newtonian acceleration due to gravity in the flat Earth approximation)
\begin{equation}
\left(\rho+{p\over c^2}\right)  g = - \nabla p.
\end{equation}

Combining, and cancelling irrelevant constants, we have
\begin{equation}
4T^4 g/c^2 = - \nabla T^4,
\end{equation}
implying
\begin{equation}
{\nabla T\over T}= -g/c^2.
\end{equation}
Integrating
\begin{equation}
T(z) = T_0 \exp(-gz/c^2).
\end{equation}

Realistically,  within the context of the flat Earth approximation we should only keep lowest-order term in the exponential, so that
\begin{equation}
T(z) = T_0 \left(1  -   gz/c^2 \right).
\end{equation}
This is exactly the Tolman effect for a radiation gas; but now derived without general relativity; by just using special relativity and Newtonian gravity. 
A refinement is to replace $g\to-\nabla\Phi$ in the relativistic Euler equation, integrating then yields
\begin{equation}
T(z) = T_0 \exp(\Phi/c^2) \approx T_0 \left(1  +\Phi/c^2 \right).
\end{equation}
Thus we see that the Tolman effect can be interpreted special relativistically, using the notions of radiation gas, $E=mc^2$, the relativistic Euler equation, and Newtonian gravity (with the observed universality of free fall being essential to side-stepping Maxwell's attempt at a no-go theorem, as discussed below).  Some readers might prefer this argument as it evades the need for the technical machinery of general relativity.

\section{Planck's blackbody spectrum}\label{S:planck}
\enlargethispage{20pt}
Tolman's argument for the radiation gas is crucially dependent on the validity of the Stefan--Boltzmann law, $\rho = a T^4$,  regardless of the presence or absence of a gravitational field.
Based on this, an interesting calculation comes from simply applying the known gravitational redshift of photons to Planck's spectral law. According to Planck, the energy density of a photon gas is given by
\begin{equation}
u = \int \frac{a\; \nu^3}{e^{h\nu/k_B T} -1} d\nu = a\, T^4.
\end{equation}
If this gas is situated in a gravitational field, each individual photon will be subjected to gravitational redshift in a way that, if $\nu_0$ is the frequency of the photon at some reference height $z=0$, we have:
\begin{equation}
\nu_0 \longrightarrow \nu = \nu_0 \sqrt{1 - 2gz/c^2}.
\end{equation}
Consequently
\begin{equation}
\label{E:P1}
u(z) = \int \frac{a\; \PC{\nu_0 \sqrt{1 - 2gz/c^2}}^3}{e^{h\nu_0 \sqrt{1 - 2gz/c^2}/k_B T} -1} d\PC{\nu_0 \sqrt{1 - 2gz/c^2}}.
\end{equation}
It is possible to directly perform the integration on equation \eqref{E:P1}, and so immediately obtain the Stefan--Boltzmann law. However, we will instead rewrite $T = T_0 \sqrt{1 - 2gz/c^2}$, which is the Newtonian approximation for Tolman's result, and thereby obtain
\begin{equation}
u(z) = \int \frac{a\; \PC{\nu_0 \sqrt{1 - 2gz/c^2}}^3}{e^{h\nu_0/k_B T_0} -1} d\PC{\nu_0 \sqrt{1 - 2gz/c^2}}.
\end{equation}
Dividing the system into horizontal slices, we can look at specific fixed heights $z$.  Doing so, we obtain: 
\begin{eqnarray}
	u(z) &=& \int \frac{a\;\nu_0^3}{e^{h\nu_0/k_B T_0} -1}  \PC{\sqrt{1 - 2gz/c^2}}^4  d\nu_0 \nonumber\\
	&=& \PC{\sqrt{1 - 2gz/c^2}}^4 \int  \frac{a\; \nu_0^3}{e^{h\nu_0/k_B T_0} -1} \;   d\nu_0 \nonumber\\
	&=& \PC{\sqrt{1 - 2gz/c^2}}^4 \; a \,T_0^4 = a\, T(z)^4.
\end{eqnarray}
This might (very naively) be seen as a circular argument, but there is an important physics point here --- demonstrating that the validity of Stefan--Boltzmann law is not affected by the presence of a temperature gradient due to gravity. 
Indeed the argument also shows that the Tolman effect can in principle be fully explained by the gravitational redshift --- which is a purely kinematic effect in any metric theory of gravity --- the Einstein equations are not used in this version of the argument leading to the Tolman effect.
In short, Tolman's result is completely consistent  with the Stefan--Boltzmann law. 

\vspace{-15pt}
\section{Defining temperature}\label{S:temperature}

Given this extended technical discussion about thermodynamics and general relativity, one might ask what precisely is the definition of temperature being used. It is indeed the standard thermodynamic definition
\begin{equation}
\label{temperature}
T(x)^{-1} = \frac{dS}{dE}.
\end{equation} 
Here $T(x)$ is the spatially dependent temperature from equation \eqref{E:tolman temp}, while $S$ is the entropy and $E$ is the energy of a small ``fluid element'' located at position $x$, as pointed out in reference~\cite{frolov}. 

An important question that might arise is this: Temperature, entropy and energy measured by whom? Given that $T(x)$ is normally referred to as ``the locally measured temperature'', the answer must be: Those are the thermodynamic quantities measured by a local observer. But what if another observer, not quite local, decides to do the same measurements? What will she see?

\enlargethispage{10pt}

Before answering that question, is it important to know how to calibrate thermometers. Given Tolman's result, $T(x) = T_0 /\sqrt{g_ {00}(x)}$, we certainly know that what each thermometer will measure will depend on its position. Then in a manner similar to clock synchronization in general relativity, we might attempt to ``synchronize'' them. But to do so, it is necessary to either set the zero of the temperature scale by placing all the thermometers at the same position (or on the same equipotential surface) or to use controlled physical processes at each height to establish the temperature there. Otherwise the temperature gradient (or lack thereof) might merely be an artefact of thermometer calibration. 

\begin{figure}[!htb]
	\includegraphics[scale=.45]{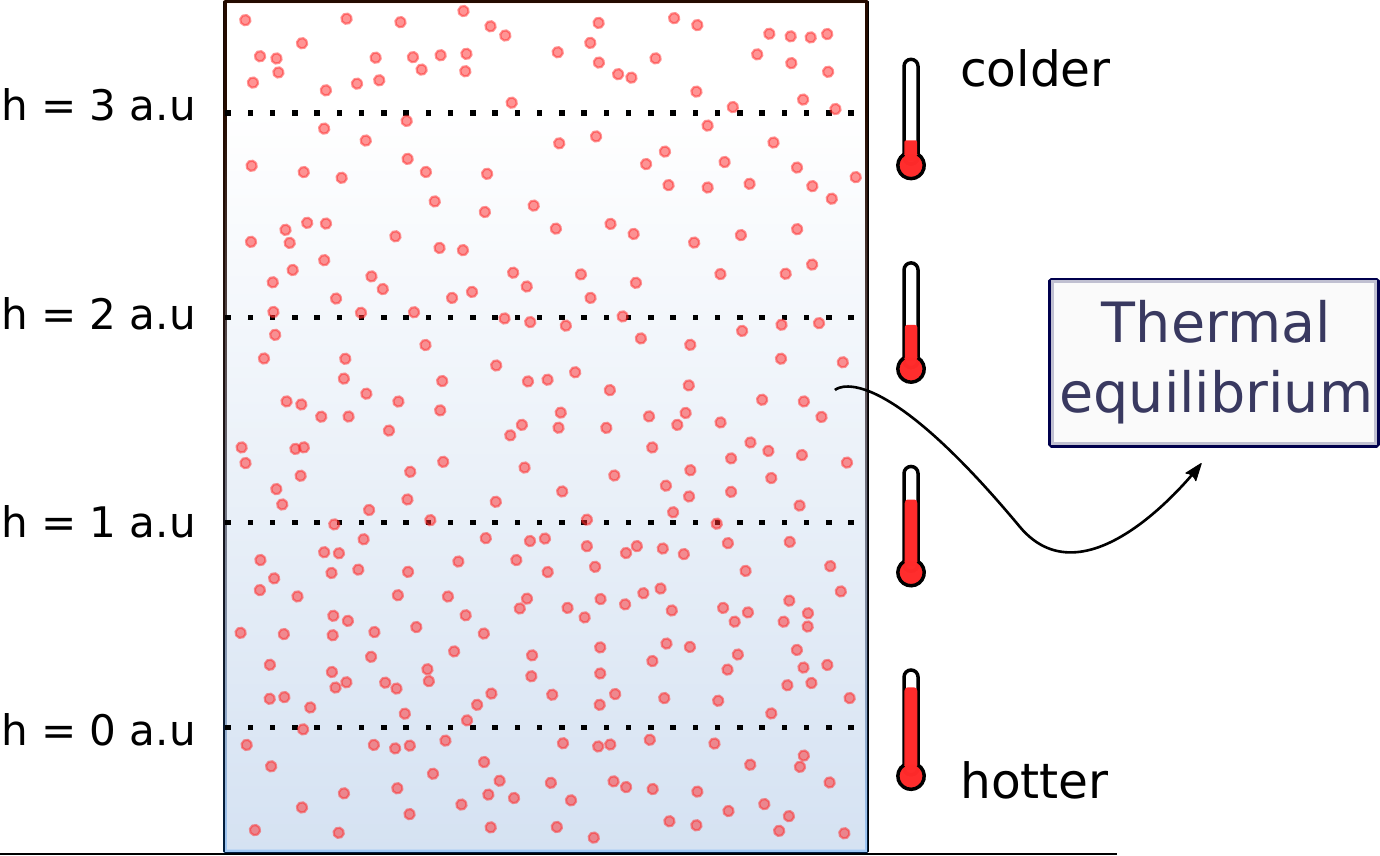}
	\caption{\label{F:thermometers}Thermometers in a gravity gradient: 
		Representative picture in arbitrary units (a.u) of the temperature gradient caused by a gravitational field.}
\end{figure}

Now, let us assume we place the carefully calibrated thermometers at different heights in a gas column, as shown in figure \ref{F:thermometers}. They will keep track of what the local observers are measuring, the position-dependent $T(x)$. However, assume also that there is an observer outside the box that wishes to know what is the internal temperature distribution of the gas without making any local measurement.

She might do that, for example, by placing some device which opens a small cavity at the desired position, in a way that a sample of the black body radiation of the gas at that height will be sent to her. 
However, in the process of travelling towards the observer, the light frequency will be modified due to gravitational redshift, which will exactly cancel the metric dependence factor in the temperature $T(x)$. 

\begin{figure}[!htb]
	\includegraphics[scale=.55]{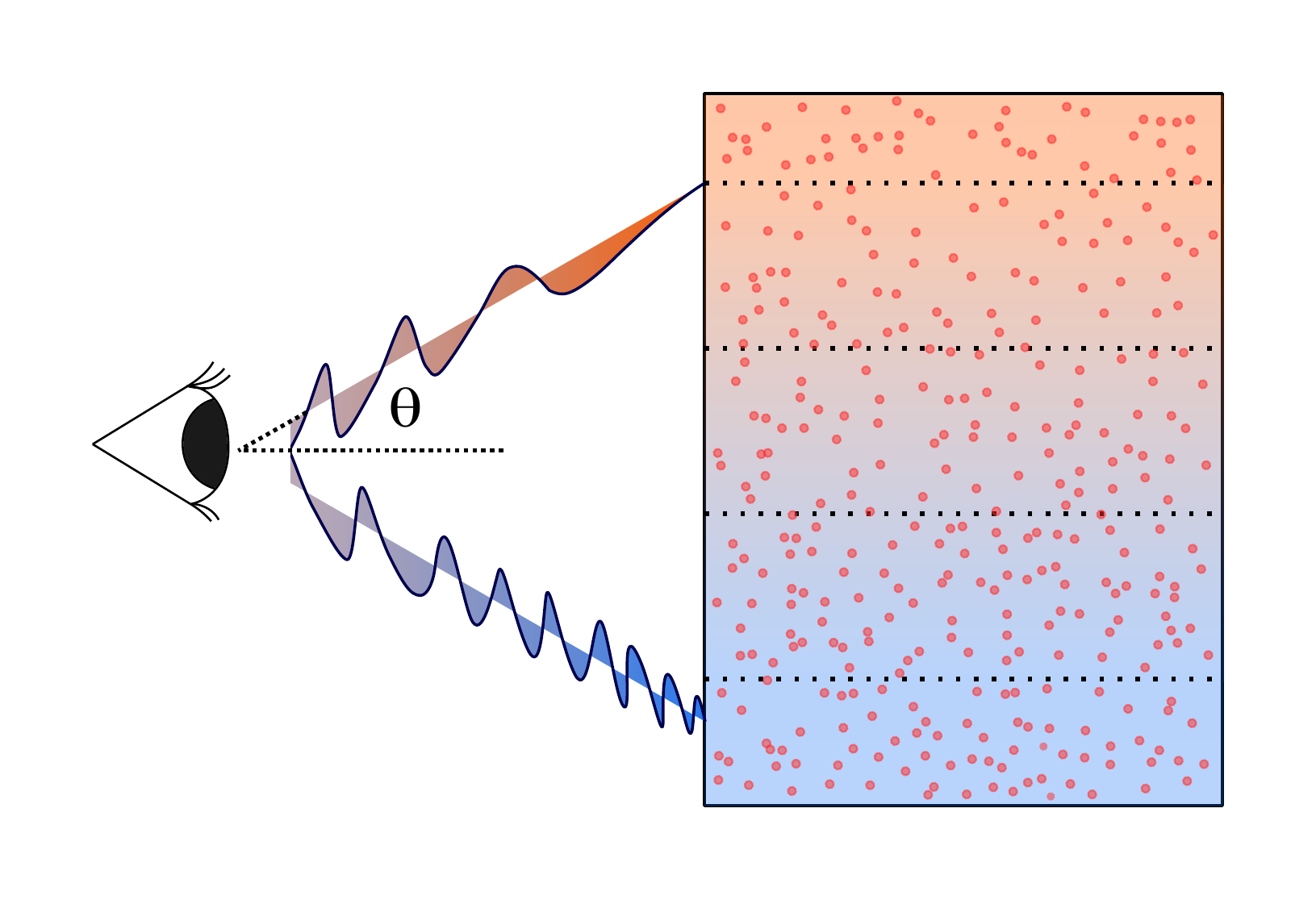}
	\caption{\label{F:thermometers2}External observer looking at photons leaking from the box containing the photon gas, 
		with the photons arriving at some angle to the vertical.}
\end{figure}

To understand this better, consider the observer to be located at $z=0$ for convenience, looking in some direction an angle $\theta$ to the horizontal (see figure~\ref{F:thermometers2}). Photons coming from a distance $r$ away from her are coming from a height $z=r \sin\theta$. Suppose, for argument's sake, the Tolman effect was \emph{not present}, (that is, if locally measured temperatures were constant), then a Planck spectrum emitted from $z=r \sin\theta$ would be redshifted/ blue\-shifted by a factor $(1+gz/c^2)$ by the time the photons arrive at the observer at $z=0$. However, light rays coming from distinct places will redshift/blueshift differently, in a way that the observer at $z=0$ would see not a simple Planck spectrum, but rather a \emph{superposition} of Planck spectra of different temperatures. But then the radiation gas is not at equilibrium at $z=0$, and we have a \emph{reductio ad absurdum}.  The only way to avoid inconsistency is if the radiation gas has a position dependent temperature $T(z)=T_0/(1+gz/c^2)$, since then the gravitational redshift guarantees that all these Planck spectra, when seen by the observer at $z=0$ will have the \emph{same} temperature $T_0$. Again, Tolman's result is completely consistent with the Stefan--Boltzmann law and the Planck spectrum for a photon gas in internal equilibrium.

As expected from the universality of free fall, the black body radiation, as seen by an external observer, will never directly ``reveal''  local accelerations in spacetime. In this way, it also becomes clear that temperature always has to be measured locally (or quasi-locally).

Another interesting point is that the constant temperature seen by the outside observer will depend on the observer's position as well, 
since they will only ``see'' the temperature that is relevant to the equipotential slice on which they are located.
That is included in the physical meaning of $T_0$ in equation \eqref{E:tolman temp}. In the constant gravity case, for example, the higher the observer's position, the smaller the measured $T_0$. In such manner, $T_0$ it is indeed a constant for each fixed external observer, but it may vary from one external observer to another. Concluding this discussion, we see that temperature, just as time, has to be measured locally or quasi-locally, even when a system is in thermal equilibrium.   

\section{Maxwell's argument}\label{S:maxwell}

We now want to look at an argument given by Maxwell~\cite{maxwell:1868} some 150 years ago, regarding the equilibrium temperature of a vertical column of gas. It is based on the second law of thermodynamics and, as we will discus, is subtly misleading when applied to gravity, although it is fully valid for other forces. Using the more recent 1902 presentation~\cite{maxwell:1902}, the first part of Maxwell's argument goes along these lines, and is certainly valid in all generality: 
	
	\begin{quote}
	 \emph{``[...]
	 if two vertical columns of different substances stand on the same perfectly conducting horizontal
	plate, the temperature of the bottom of each column will be
	the same; and if each column is in thermal equilibrium of
	itself, the temperatures at all equal heights must be the same. 
	In fact, if the temperatures of the tops of the two columns
	were different, we might drive an engine with this difference of temperature, 
	and the refuse heat would pass down the colder column, through the conducting plate, and up the warmer
	column; and this would go on till all the heat was converted
	into work, contrary to the second law of thermodynamics.''
	}
	\end{quote}
	
This first part of Maxwell's 1902 argument establishes that temperature gradients in equilibrium states, if present at all, must be \emph{universal}, otherwise the Clausius version of the second law is violated. (Temperature differences at the \emph{same height} will certainly drive heat fluxes, and would allow one to construct a \emph{perpetuum mobile}.) 
	
Now this is not exactly what Maxwell originally concluded, because he was primarily interested in non-relativistic atomic and molecular gases. The second part of his original argument went as follows:


	\begin{quote}
	\emph{``But we know that if one of the columns is gaseous, 
	its temperature is uniform \emph{[from the kinetic theory of gases]}.    
	Hence that of the other must be
	uniform, whatever its material.''
	}
         \end{quote}
This second part of Maxwell's argument is now known to be incomplete once one includes relativistic effects. 

As we have seen, to obtain his \emph{reductio ad absurdum} result Maxwell made two quite specific assumptions: 1) that the (non-relativistic) kinetic theory result regarding the temperature of vertical gas column is true, so gases have zero temperature gradient when in thermal equilibrium regardless of the presence or absence of gravity, and, 2) that the temperature gradient, if exists, is different for distinct substances. These two strong assumptions, when put together, indeed leave not enough space for evading a \emph{perpetuum mobile}.

Another possible version of this argument, which does not use the kinetic theory result \emph{a priori}, but keeps the substance dependence assumption, can be formulated as follows: Assume you have a vertical column of gas in a gravitational field and suppose that, after equilibrium is reached, a vertical temperature gradient is present. If this is true, we can use a wire or some other heat permeable material to connect the upper and lower parts of the gas container and create, just like in Maxwell's scheme, a \emph{perpetuum mobile} of the second kind. 

The reason this second argument is again misleading is based on the universality of general relativity, which translates to the statement that any form of mass or energy is equally subjected to gravity. With the development of general relativity we became aware that gravity does not concern forces between bodies. It is about space-time, curvatures and geodesics. So, it doesn't matter whether we are looking at a gas, a piece of lead or photons. They will all experience the same metric and the effects that arise from it. 

In this way, we see that if we use a wire to connect the top and the bottom of the gas container, all the atoms comprising the wire will also be suffering gravity's influence, in exactly the same way as the atoms in the gas. So the wire itself will exhibit a vertical temperature gradient, which is exactly the same as that in the gas, making the idea of a thermal machine impossible, since all its components would be in thermal equilibrium at every individual horizontal slice. The same argument is valid for Maxwell's two-column system. 

It is worth noting that the speed of light explicitly occurs in Tolman's temperature gradient $(\nabla T)/T = - g/c^2$, and so this effect would not and could not be picked up in any non-relativistic analysis.   
Maxwell's claim of the absence of temperature gradients in gaseous equilibrium is certainly an excellent approximation in the non-relativistic regime; it is only in relativistic physics that (tiny) equilibrium temperature gradients show up. 

To conclude, it is important to point out that Maxwell's argument is only \emph{evaded} due to gravity's universality. In that fashion, one might still possibly apply Maxwell's argument  to other forces, as we will do in the following section.

\vspace{-10pt}
\section{Electric thermal gradient?}\label{S:electric}
\enlargethispage{20pt}

While trying to understand the theoretical underpinnings of  Tolman's gravitational temperature gradient, we might ask ourselves whether it is an effect that could also be generated by any external potential that breaks isotropy and homogeneity of space, or whether it is specific to general relativity (possibly special relativity) and its many peculiar features. 
To clarify this point, we will consider an electric analogue of the gas column in a gravitational field, and analyse some consequences that an electrically induced thermal gradient would create. From them, we will be able to infer something about the plausibility of an electric temperature gradient.  (Spoiler alert: No, it is not plausible.)

Consider an electron gas inside a box. An external electric field will be assumed to act on the whole system for long enough so that the particles already have had sufficient time to rearrange themselves into a equilibrium situation. Assume also that the gas density is very low, so that the force exerted by the external field is much stronger than the interactions between individual electrons
(although they \emph{do} interact in order that thermal equilibrium be achieved).
If any temperature gradient occurs, it will be aligned with the  direction of the external electric field. For simplicity, assume no gravitational field is present.

Let us now (for the sake of argument) assume that a temperature gradient in the equilibrium configuration does exist and ask what the possible thermodynamic consequences might be? A possible way to answer that question is to take the same path that Maxwell's argument followed. To do so, let us assume that we place, next to the just mentioned electron gas, a box filled with electrically neutral particles, \emph{i.e.}, photons, neutrons, \emph{etc.}
Due to its neutrality, this second box will not interact with the electric field, thus having no reason at all to present a temperature gradient. Continuing the argument on the same lines as before, we might use the gas of neutral particles as a ``conducting wire'', connecting the two ends (of the electron gas box) with distinct temperatures. This will create a heat flow,  which enables us to construct a perpetual motion machine of the second kind. 
This argument is valid since, unlike gravity, electric fields are \emph{not universal}, given that the effect it will cause on a particle depends on the particle's electric charge. 
%
\begin{center}
\begin{figure*}[!htb]
	\includegraphics[width=.9\textwidth]{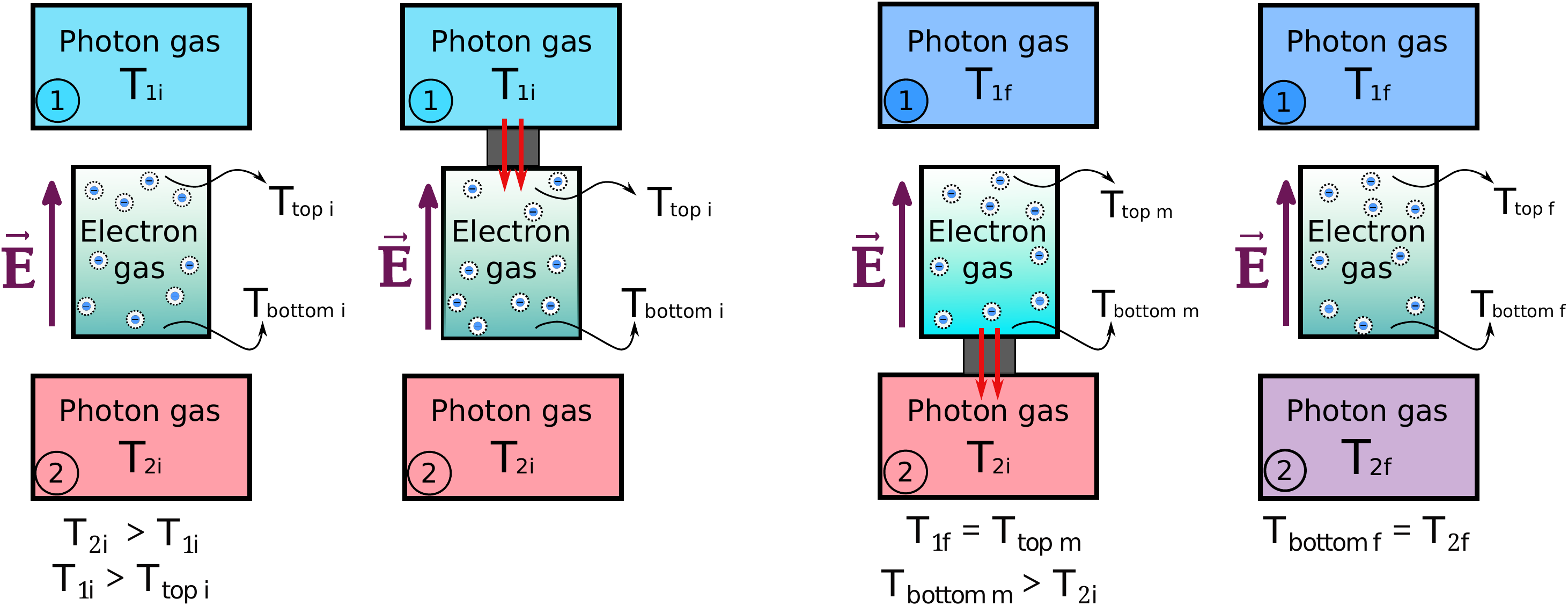}
	\caption{\emph{Gedankenexperiment:} Heat engine showing how heat is being transferred from the cold to the hot photon gas. Since heat flows depend only on the quasi-local distributions of temperature, it is possible to transfer heat from Box 1 to the electron gas, followed by a heat transfer from the electron gas to Box 2. In the final stage we have $T_{1f} < T_{1i}$ and $T_{2f} > T_{2i}$, which violates the second law of thermodynamics.}
	\label{F:heat}
\end{figure*}
\end{center}
%
\null
\vspace{-40pt}
For the sake of clarity, we will also explicitly show that, if electric fields are able to produce temperature gradients in a gas in thermal equilibrium, then heat engines that violate the second law can be easily created. We will use a 
``\emph{gedankenexperiment}'' to do so.
In the system presented in figure~\ref{F:heat} we have three boxes aligned in the direction of an external constant electric field (vertically). The boxes labelled 1 and 2 contain radiation gas while the middle container is filled with an electron gas. As the external electric field is applied everywhere, if it can indeed create temperature gradients, the temperatures at the top and at the bottom of the electron gas will be such that $T_{top\; i} > T_{bottom\; i}$. The temperatures of the photon gases are constant. 

Now we will choose the temperatures of the boxes wisely. Box 1 will be colder than 2, but it will be hotter than the top temperature of the electron box: $T_{2 i} > T_{1 i} > T_{top\; i}$. In this way, if we connect  Box 1 with the electron gas, the laws of thermodynamics tell us that heat will flow to the latter  until the top temperature of the electron gas equalizes with Box 1's temperature. The only assumption we are making here is that heat transfer depend on the local temperatures where the boxes touch. So, although the electron gas has (by assumption) a position-dependent temperature, it is the temperature where the contact is made with the photon gas that will tell us whether a heat flow will occur or not. 

After this step, Box 1 will be colder and the electron gas warmer than its initial state, with $T_{1 f} = T_{top\; m} > T_{top\; i}$, and $T_{bottom\; m} > T_{bottom\; i}$ after equilibrium is reached. Additionally, we demand that the temperature of Box 2 be such that after the first heat transfer, $T_{bottom\; m} > T_{2 i}$. In this way, if we now connect Box 2 with the electron gas, given the temperature differences, heat will flow to Box 2 until its temperature equalizes with the bottom temperature of the electron gas. In the final picture we have temperatures satisfying $T_{1f} < T_{1i}$ and $T_{2f} > T_{2i}$. The final average temperature of the electron gas will depend on the heat capacity of both photon boxes.

But this means that heat was transferred from a cold to a hot body, without any work being done on or by the system, which is a clear violation of the second law of thermodynamics. As the construction of the argument is extremely simple and depends only on the non-universal character of the electric force, it is easy to extend it to any force that is not universal.

We might state the conclusion of this argument as: 

\emph{Given that temperature gradients created by any force that is not universal (e.g. dependent of charge, mass, spin,...) allows the creation of heat machines that violate the second law, these temperature gradients must not exist. }

Going even further, up to date no force other than gravity seems to act on all sources of mass or energy in the same way regardless of internal composition of the bodies, so if desired we might even state this as:

\emph{Gravity is the only force capable of creating temperature gradients in thermal equilibrium states without violating the laws of thermodynamics.}

\vspace{-15pt}
\section{Conclusions}\label{S:conclusions}

In this article we have discussed some physical consequences that arise from Tolman's spatial dependence of locally measured equilibrium temperature distributions in gravitational fields --- the best framework for doing this analysis being general relativity. 
(Though special relativity, by combining the notions of photon gas, $E=mc^2$, the special relativistic Euler equation, and Newtonian gravity is enough to capture the basic physics.) We have argued that  temperature, just as time, needs to be measured locally (or at worst quasi-locally). In addition, when dealing with the black-body radiation emitted by a finite-size system, we have seen that the frequency spectrum observed by an external observer will be blueshifted/redshifted  by an amount exactly cancelling Tolman's $\sqrt{g_{00}}$ factor, in such a way that observers looking at a finite-size body will ``perceive'' the same temperature at all points in the body. In addition, we have also pointed out how this observed temperature will depend on the external observer's position.

\bigskip

Next, an electric analogue of the ideal gas in a static gravitational field was presented. Based on the thermodynamic consequences that an electrically induced thermal gradient would have, it became clear that such an effect \emph{cannot} exist 
(this is effectively an extension of Maxwell's no-go argument).
Finally, a further natural extension of such ideas leads to the conclusion that gravity is the \emph{only} force capable of creating temperature gradients in thermal equilibrium states without violating the laws of thermodynamics.

In closing, we should emphasize that while the specific points raised in this article are largely novel, the existence of Tolman's temperature gradient is not at all controversial (at least not  within the general relativity community). See for instance references~\cite{Israel:1976,Abreu:2010a,Abreu:2010b,Abreu:2010c,Abreu:2011,Padmanabhan:2003,Padmanabhan:2010,Padmanabhan:2017,Haggard:2013}.

\vspace{-20pt}
\acknowledgements
JS was supported by a Victoria University of Wellington PhD Scholarship.
JS wishes to thank Cesar Uliana-Lima for helpful discussions.\\
MV was supported by the Marsden Fund, administered by the Royal Society of New Zealand. 
\appendix
\vspace{-20pt}
\section{Counterfactual history}\label{S:counterfactual}

\vspace{-10pt}
As early as 1873 (in untranslated work by Nikolay Umov largely unknown in the West) there were suggestions that $E=kmc^2$, for some unspecified constant $k$~\cite{umov}. 
This should immediately have suggested a mass density $\rho = k^{-1} u/c^2$ for radiation gas. The \emph{ordinary non-relativistic} Euler equation, $\rho g = -\nabla p$, would then yield
\begin{equation}
  {T^4  g\over k c^2} = -{1\over3} \nabla T^4.
\end{equation}
Integrating, this leads to
\begin{equation}
T(z) = T_0 \exp\left( - {3gz\over 4kc^2}   \right) \approx T_0 \left( 1 - {3gz\over 4kc^2}   \right).
\end{equation}
While the precise value of this temperature gradient does differ from Tolman's result by the purely numerical order-unity factor of $3/(4k)$, 
all of the really essential physics was already implicitly there --- a Tolman-like result could have been in place a quarter of a century before 1900. 

\section{Loschmidt's argument}\label{S:loschmidt }

In the mid-late 1800s there was an extended dispute between Loschmidt and Maxwell/Boltzmann regarding possible gravitationally induced temperature gradients in non-relativistic gases. To explain the nature of the controversy, let us first consider an extremely ``thin'' monatomic gas, where the mean-free-path is very much larger than the size of the container. (So the only atomic collisions are elastic collisions with the walls of the container.) Under these conditions, each individual atom certainly satisfies
\begin{equation}
v(z)^2 = v(0)^2 - 2 g z,
\label{E:free}
\end{equation}
and averaging over all atoms one certainly has
\begin{equation}
\langle v^2\rangle_z = \langle v^2\rangle_0 - 2 g z.
\end{equation}
Define an effective non-equilibrium ``temperature'' by
\begin{equation}
 m \langle v^2\rangle_z =  3 k_B T_{\not=}(z). 
\end{equation}
(Since the gas is assumed ``thin'', there are as yet no internal collisions to drive it into equilibrium. So this is not a temperature in the normal thermodynamic sense. It does however match the usual definition of temperature in terms of average kinetic energy.)
Nevertheless, even if this is not a physical temperature, one can then formally write
\begin{equation}
\label{E:L1}
T_{\not=}(z)  = T_{\not=}(0)  \left\{  1 - {gz \over {1\over 2}  \langle v^2\rangle_0 } \right\}.
\end{equation}

This at first glance appears similar to Tolman's result, but 
Tolman's result is smaller than this by a factor
\begin{equation}
\hbox{(thermal velocity)}^2/\hbox{(speed of light)}^2 \approx 10^{-12}.     
\end{equation}
To better see what is going on, write this as 
\begin{equation}
\label{E:L2}
T_{\not=}(z)  = T_{\not=}(0)   -  {2m gz \over 3 k_B}.
\end{equation}
Note that this is \emph{not} universal, explicitly  depending on the atomic mass.
As long as the gas is extremely ``thin'' this is however a perfectly legitimate (if perversely presented) result. It side-steps Tolman's arguments because the ``gas'' is not a fluid,  
one is merely considering an ensemble of non-interacting particles. It side-steps Maxwell's thermodynamic arguments because the ``gas'' is not in internal thermal equilibrium.

However, if the gas becomes ``thick'' (the mean free path becomes smaller than the size of the container) the physics radically changes --- first of all a formula along the lines of \eref{E:free} only holds \emph{between} interatomic collisions,
\begin{equation}
v(z_f)^2 = v(z_i)^2 - 2 g (z_f-z_i).
\label{E:free2}
\end{equation}
Interatomic collisions, while elastic, now allow for the transfer of energy between different horizontal layers in the gas. 
What is conserved in atomic collisions is
\begin{equation}
(v_1^2+v_2^2)_\mathrm{before} = (v_1^2+v_2^2)_\mathrm{after}.
\end{equation}
This allows kinetic energies to diffuse throughout the gas column.
Once internal equilibrium is reached, one has $T_{\not=} \to T$, which is now the actual physical temperature based on kinetic theory.  Maxwell's argument then applies, as do the various other arguments adduced in this article, and equations \eref{E:L1} and \eref{E:L2} cannot survive, simply because they are not universal. 

Maxwell and Boltzmann were working in the non-relativistic approximation (where formally $c\to\infty$) the kinetic theory of gases then yields
\begin{equation}
T(z) = T_0.
\end{equation}
Instead, icluding relativistic effects (keeping $c$ finite) one has Tolman's result
\begin{equation}
T(z) = T_0 \left( 1 - {gz\over c^2} \right). 
\end{equation}
Either way, equations \eref{E:L1} and \eref{E:L2} simply do not survive once interatomic collisions are included in the discussion.






\begin{thebibliography}{69}
\bibitem{footnote:0}
Somewhat confusingly, in Tolman's original articles, he uses $T_0$ to denote the locally measured position-dependent temperature --- which we denote $T(x)$. In contrast, we reserve $T_0$ for the redshifted temperature, which is spatially constant in thermal equilibrium.
\bibitem{footnote:1}
The use of isotropic spatial coordinates is merely a convenience, it is not essential to the argument.
\end{thebibliography}
\end{document}